\documentclass[11pt,twoside]{article}
\usepackage{asp2010}

\resetcounters

\begin{document}

\title{Star-Planet Interactions in X-rays}
\author{Katja Poppenhaeger$^1$
\affil{$^1$Hamburger Sternwarte, Gojenbergsweg 112, 21029 Hamburg, Germany}}

\begin{abstract}
We investigated the coronal activity of planet-hosting stars by means of statistical analysis for a complete sample of stars in the solar neighborhood. We find no observational evidence that Star-Planet Interactions are at work in this sample, at least not at the sensitivity levels of our observations. We additionally test the $\upsilon$~Andromedae system, an F8V star with a Hot Jupiter and two other known planets, for signatures of Star-Planet Interactions in the chromosphere, but only detect variability with the stellar rotation period.
\end{abstract}

\section{Introduction}
Interactions between stars and close­-in planets can be expected from the analogy to binary stars. Binaries are often more active than single stars of the same spectral class \citep{AyresLinsky1980}, and X-­ray emission between the two components of a binary has been observed as well \citep{Siarkowski1996}. Thus, regarding stars with giant planets as binaries with an extremely small mass ratio, one expects to see enhanced activity levels of the host star from tidal or magnetic interaction with the planet \citep{CuntzSaar2000}, which should manifest themselves in activity proxies such as chromospheric Ca~II emission and coronal X-­ray emission. If Star­-Planet-­Interactions (SPI) are observed reliably, they can yield valuable information on the magnetic fields of exoplanets, the irradiation of exoplanetary atmospheres by the host star which in turn affects planetary evaporation \citep{Vidal-Madjar2003}, as well as orbital synchronization and planetary migration timescales.

\section{Analysis of the stellar sample}
\subsection{Stellar sample: data analysis}
We constructed a sample of all planet-hosting stars within $30$~pc distance from the Sun as known at the time of analysis, summing up to a total of $72$ stars \citep{Poppenhaeger2010}. For some of these, X-ray properties were known from previous ROSAT or XMM-Newton observations, but for a large number of these stars X-ray characteristics were not or only poorly known. Therefore we observed 20 planet-hosting stars with XMM-Newton to determine X-ray luminosities for stars which had not been detected before in other X-ray missions, and to derive coronal properties from spectra recorded with EPIC and, given sufficient signal, RGS detectors, especially for stars with close-in planets. We reduced the data with SAS version 8.0, using standard criteria for filtering the data. We extracted counts from the expected source regions with radii between $10\arcsec$ and $30\arcsec$, depending on the source signal, background conditions and the presence of other nearby sources. Background counts were extracted from much larger, source-free areas on the same chip for the MOS detectors and at comparable distances from the horizontal chip axis for the PN detector.

We determined the corresponding X-ray luminosities via calculating count-to-energy conversion factors (CECFs) for several energy bands. For stars with sufficient numbers of accumulated X-ray photons, we verified the results alternatively by spectral fitting in Xspec~v12.5, finding very good agreement between both methods. For the error estimate on the luminosities we used Poissonian errors on the count numbers and added an additional uncertainty of $30\%$ of the total luminosity to account for intrinsic stellar variability and errors in the CECFs.

The two X-ray missions from which we use data here, XMM-Newton and ROSAT, have different X-ray energy bands to which they are sensitive. We therefore normalized the XMM-Newton luminosities to the ROSAT band, using different normalization factors depending on the hardness ratios of the individual XMM-Newton sources.

\begin{figure}[t!]
\plottwo{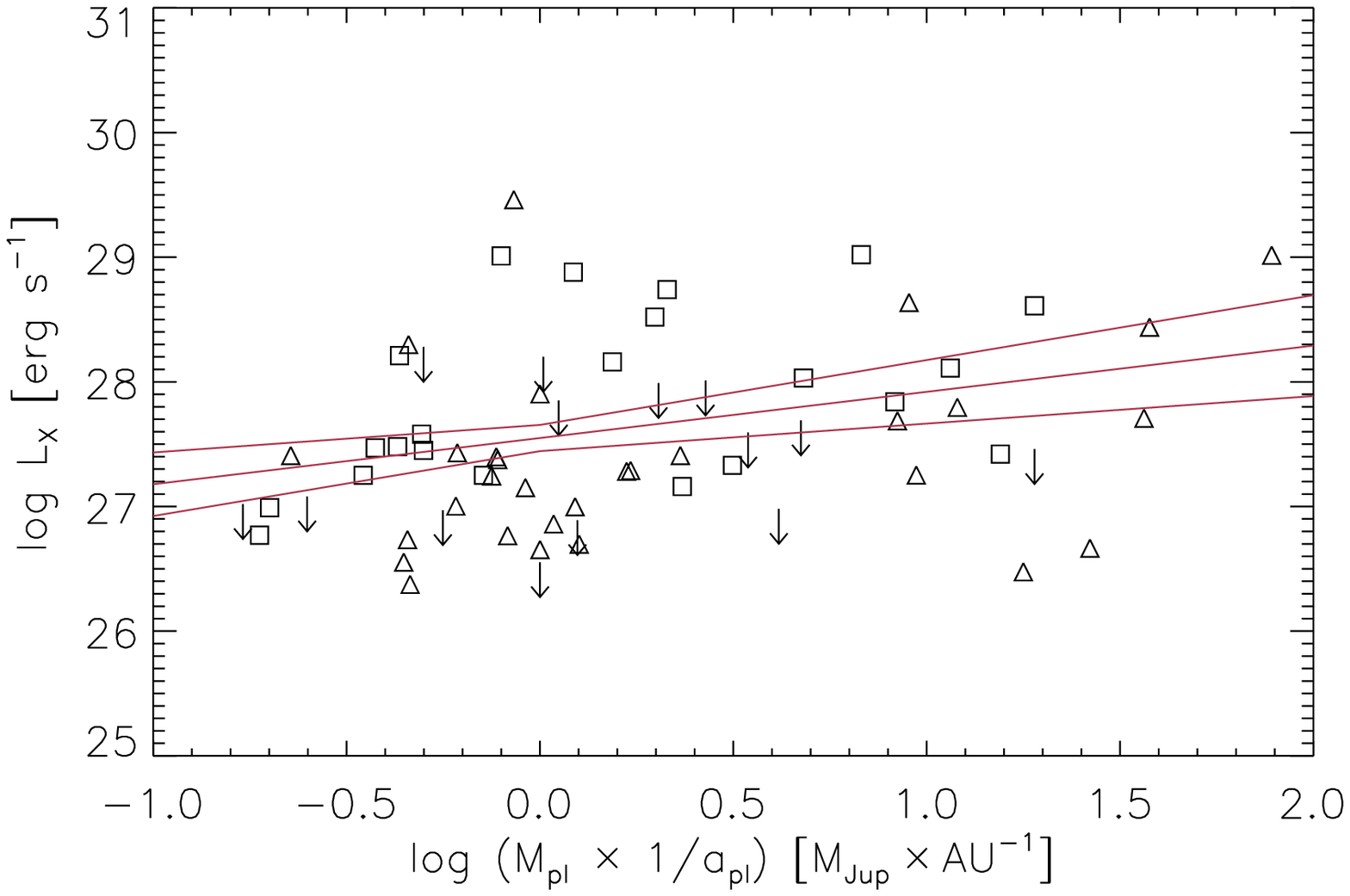}{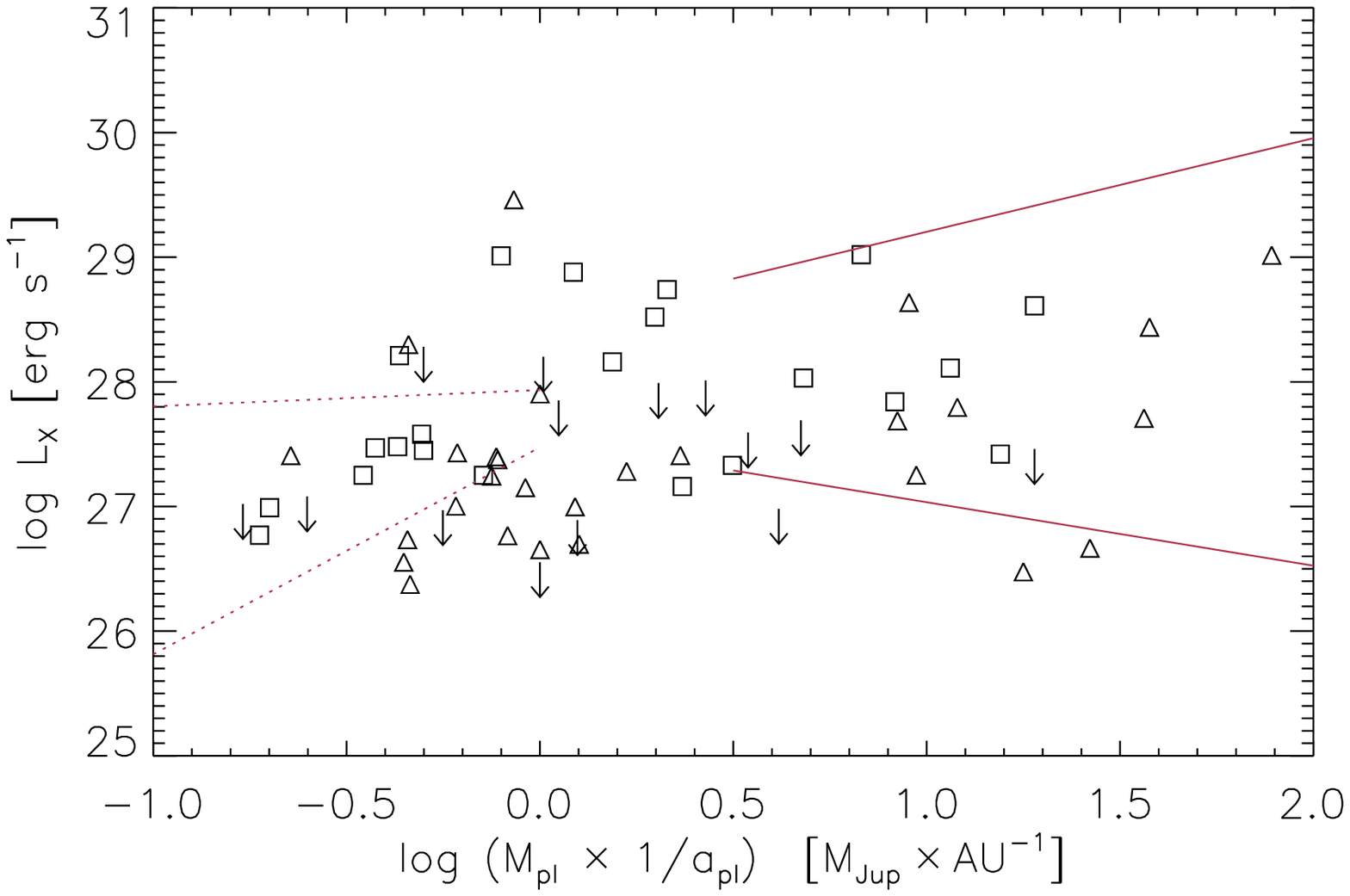}
\caption{{\it left panel:} X-ray luminosity versus planetary mass and inverse semimajor axis for the investigated stellar sample; XMM-Newton detections plotted as triangels, ROSAT detections as squares. {\it right panel:} separate regression analyses performed for the outer subsamples. Both trends overlap well within errors, indicative of no SPI-related trend in this sample.}
\label{lx}
\end{figure}

\subsection{Stellar sample: results}

We tested the complete set of data for correlations between planetary parameters (mass $M_{pl}$ and semimajor axis $a_{pl}$) and stellar X-ray properties (X-ray luminosity $L_X$ and activity indicator $L_X/L_{bol}$) using Spearman's $\rho$ rank correlation coefficient. The only significant correlation which is present in the data is a correlation of $L_X$ with the product of planetary mass and inverse semimajor axis $M_{pl}\times a_{pl}^{-1}$. The correlation coefficient yields $\rho = 0.31$, corresponding to a probability of $3\%$ that this value can be reached randomly.

To visualize this correlation, we plot the logarithmic X-ray luminosity versus the logarithm of planetary mass times inverse semimajor axis (see Fig.~\ref{lx}, left panel). The key question is: is this trend induced by X-ray signatures of SPI, or is it due to possible selection effects?

For all but three stars in our sample, the planets were detected by the radial velocity (RV) method. Stellar activity masks the RV signal, so around active stars, only massive, close-in planets can be detected since they induce a strong RV signal, while low activity of a star makes the detection of planets with lower mass or larger semimajor axis easier. Indeed, the star for which the first planet was detected (also using the RV method), 51~Peg, shows very low activity on timescales of several years and might actually be in a Maunder minimum state \citep{Poppenhaeger2009}. 

So, the RV selection effect produces a trend which is similar to the one we detect in our data. To check if there is an {\it additional} trend present on top of this bias, we do the following analysis: we conduct a linear regression of $\log L_X$ versus $\log (M_{pl}\times a_{pl}^{-1})$ for two subsamples described by $\log (M_{pl}\times a_{pl}^{-1}) > 0.5$ (heavy, close-in planets) and $\log (M_{pl}\times a_{pl}^{-1}) < 0$ (small, far-out planets). The trend in the far-out sample should be dominated by the RV selection effect, since SPI is expected to be at work only for small distances. The close-in sample should show the RV trend plus a potential trend from SPI. The result is shown in Fig.~\ref{lx}, right panel. The two trends overlap well within their statistical errors; no {\it additional} activity trend which might be induced by SPI is detectable.

\section{Analysis of the $\upsilon$~And system}

The $\upsilon$~And system consists of a main-sequence star of spectral type F8 and three planets, all of them detected by the radial velocity method. The innermost planet has a mass of $0.69$~M$_{Jup}$ and an orbital period of $4.62$~d, corresponding to a semimajor axis of $0.059$~AU. For this system, \citet{Shkolnik2005} found hints in chromospheric data that the activity of the host star changes in phase with the planetary orbital period, indicating magnetic SPI being possibly at work. In subsequent observations however, the star showed variability mainly with the stellar rotation period \citep{Shkolnik2008}.

\subsection{$\upsilon$~And: data analysis}
To investigate the time-dependent behaviour of $\upsilon$~And's chromospheric activity, we collected $23$ optical spectra with the FOCES echelle spectrograph at Calar Alto Observatory in Spain. The coronal activity was also investigated with Chandra ACIS-S and is described in \citet{PoppenhaegerLenz2010}. The optical spectra were flatfielded and wavelength-calibrated with Thorium-Argon frames taken during the same night as the individual science frame; most of the observations were performed in July and September/October 2009. For the spectra from July 2009, the FOCES $15\mu$ detector was used, while the subsequent observations used a $24\mu$ detector. We normalized the spectra with respect to each other, paying specific attention to the Ca~II K line region, since Ca~II K (and, on a somewhat lower scale, Ca~II H) line emission is a strong indicator for chromospheric activity. We then computed a median spectrum for each of the two data groups and calculated the residuals of each individual spectrum with respect to the corresponding median spectrum. 

\subsection{$\upsilon$~And: results}
\begin{figure}[t!]
\plottwo{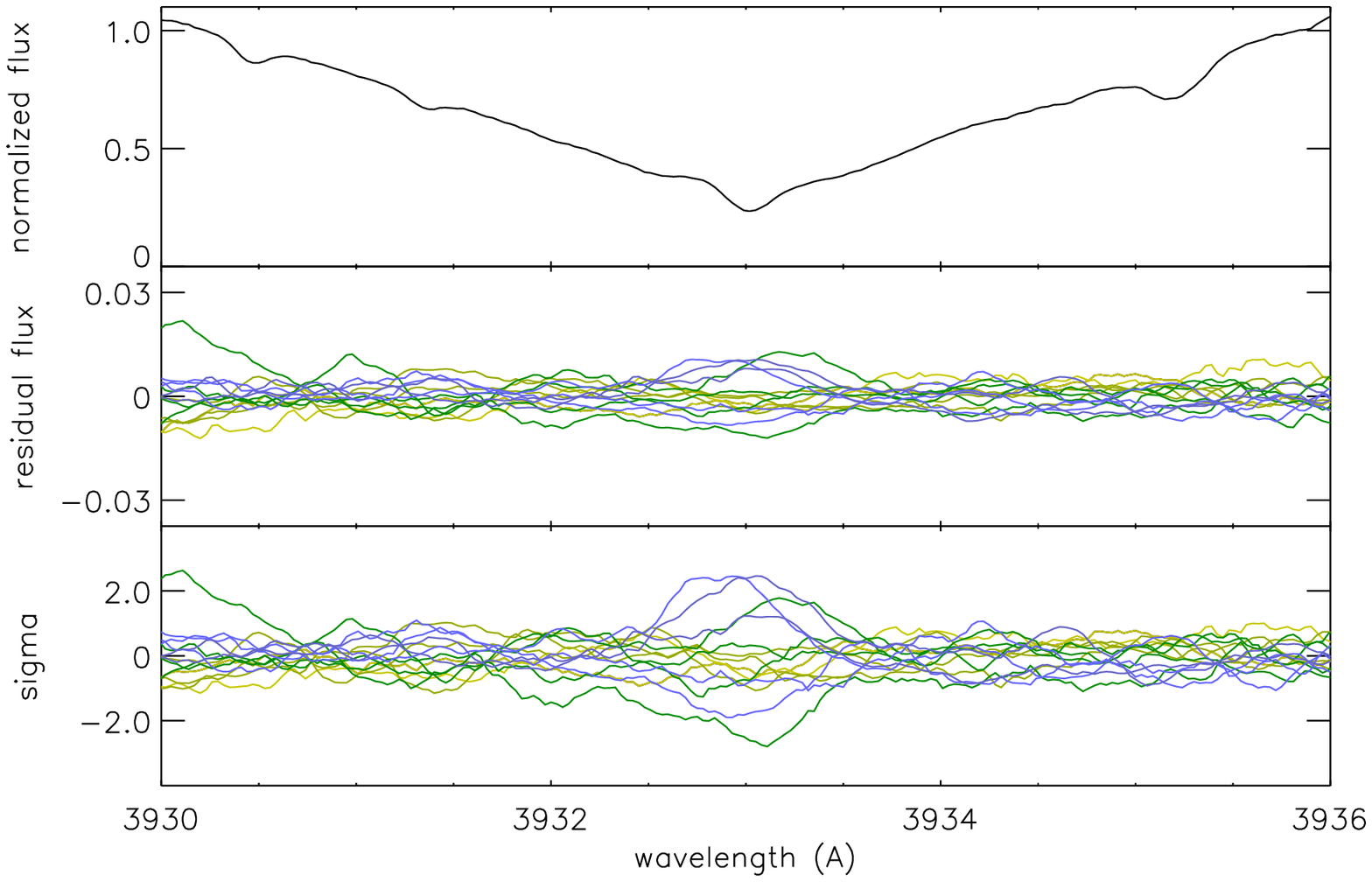}{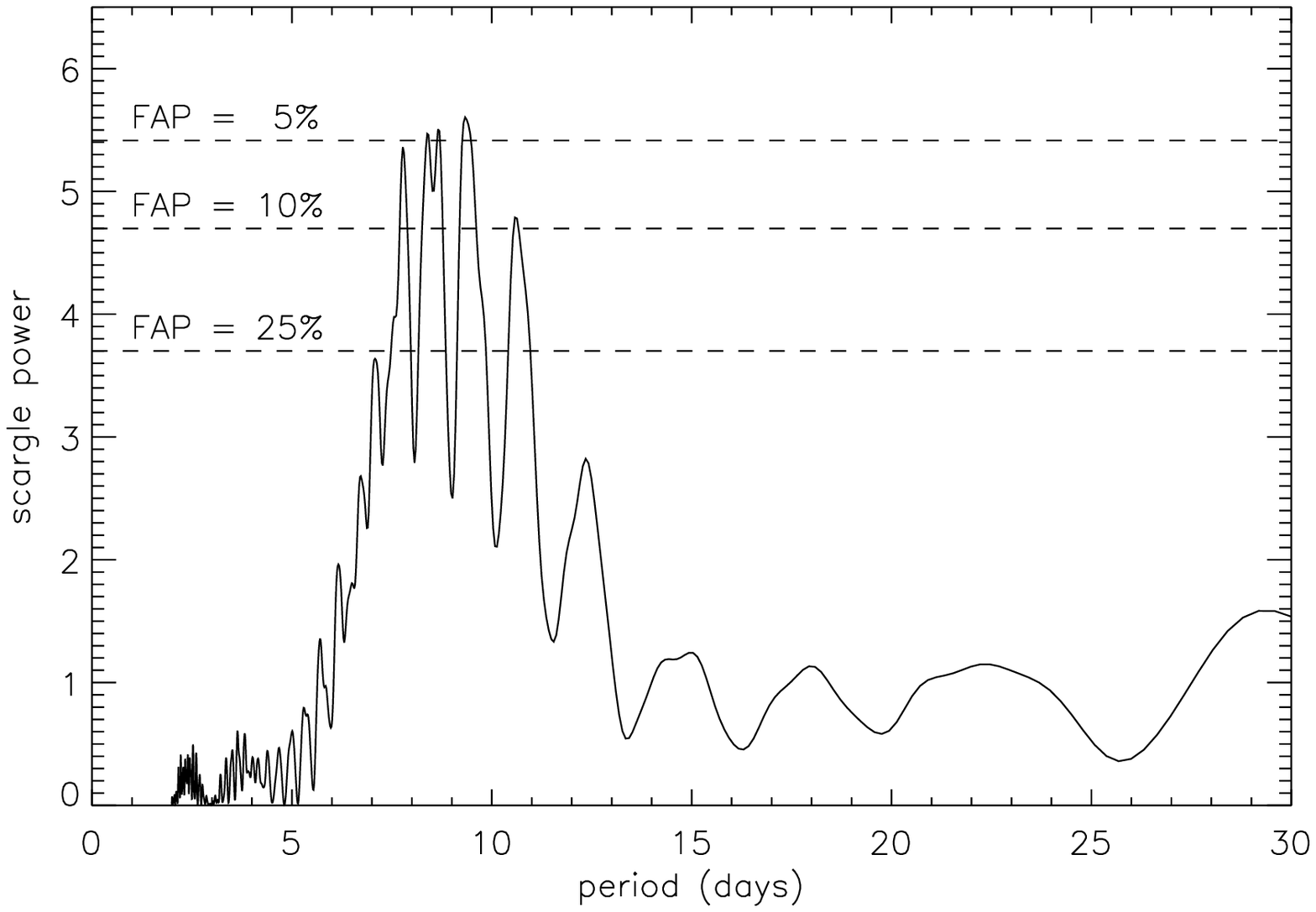}
\caption{{\it left panel:} Variability in Ca~II K line cores of $\upsilon$~And in the FOCES 15μ data.
upper part: normalized mean spectrum; middle part: residual
flux in the same normalization; lower part: flux variation in
standard deviations. {\it right panel:} Lomb-Scargle periodogram of Ca~II K line residuals
(weighted by their respective errors) with false alarm
probabilities given by the horizontal lines.}
\label{ca_res}
\end{figure}

While the overall stellar activity of $\upsilon$~And is low and the K line emission is weak, the residuals in the Ca~II K line cores indeed show some variability (see also \citet{PoppenhaegerLenz2010}). This is shown in Fig.~\ref{ca_res} (left panel) for the $15\mu$ data, which was recorded under very favorable weather conditions. The upper part of the figure shows a close-up of the Ca~II K line core; the middle part shows the residuals, smoothed by $15$~pixels; the lower part shows the variation of each $15\mu$ spectrum measured in standard deviations. For the $24\mu$ data, the noise level is higher and there is no deviation in the K line core discernible by naked-eye inspection.

We therefore integrated the residuals over the width of the line core ($1$~Angstrom) and calculated a Lomb-Scargle periodogram of the data, weighted by their respective errors \citep{Lomb1976, Scargle1982, GillilandBaliunas1987}. If the stellar activity is dominated by SPI effects, the main period that should show up is the planetary orbital period of $4.6$~d. However, we find significant peaks in the periodogram for periods of $8.2$~d, $8.7$~d and $9.3$~d (see Fig.~\ref{ca_res}, right panel). The highest peak at $9.3$~d is close to twice the orbital period. However, our July observations consist of $14$ successive nightly spectra which follow a $\approx 9$~d sinusoidal variation closely and make it unlikely that we see an alias of the orbital period, but not the orbital period itself. The periodicity might fit the stellar rotation period; the values given in the literature vary quite a lot. \citet{WrightMarcy2004} give a rotational period of $12$~d from spectroscopic monitoring; \citet{HenryBaliunas2000} find only weak signatures of rotational modulation with periods of $11$~d and $19$~d respectively in two different data sets. The rotational period calculated from the measured rotational velocity of $v\sin i = 9.5\pm 0.4$~km/s \citep{Gonzales2010} and the modelled stellar radius given as $1.6$~R$_{\sun}$ in \citet{HenryBaliunas2000} yields $\approx8.5$~d. \citet{HenryBaliunas2000} also state that the difference to the estimate derived from $v\sin i$ measurements might be due to differential rotation. We interpret our findings as observational signatures of stellar rotation; continued monitoring will help deriving a more precisely determined rotation period for $\upsilon$~And. If SPI signatures are present in this system, they are weaker than the intrinsic stellar variability during our observations.

\section{Summary}
Our investigations show that possible Star-Planet Interactions do not have a major influence on the average X-ray luminosity or $L_X/L_{bol}$ in nearby stars, at least not at the given sensitivity levels of our observations. Also our measurements of the chromospheric activity of the promising star-planet system $\upsilon$~And show no indications for SPI, but rather variability with the stellar rotation period. SPI seems to induce only small effects on the activity of the host stars; if observed over longer timescales and for more targets, however, they can provide insight into planetary and stellar magnetic fields.

\acknowledgements K.~P.~acknowledges financial support from DLR grant number 50OR0703. This work is based on observations obtained with the {\em XMM-Newton} and {\em ROSAT} X-ray missions and the {\em FOCES} echelle spectrograph at Calar Alto Observatory.

\bibliography{poppenhaeger_k}

\end{document}